\documentclass[twocolumn,aps,prb,floatfix,nobalancelastpage]{revtex4}

\usepackage[dvips]{graphicx}
\usepackage{dcolumn}
\usepackage{bm}

\begin{document}

\title{Dynamic nuclear polarization induced by hot electrons}

\author{Yosuke Komori}
 \email{komori@dolphin.phys.s.u-tokyo.ac.jp}
\author{Tohru Okamoto}

\affiliation{Department of Physics, University of Tokyo, Hongo, Bunkyo-ku, Tokyo 113-0033, Japan}

\begin{abstract}
A new method for local dynamic nuclear polarization is demonstrated in a GaAs/AlGaAs heterostructure at the Landau level filling factor $\nu=3$.
Using a narrow channel sample, where the width varies stepwise along the electron flow, we find that electron cooling (heating) causes the polarization of nuclear spins against (toward) the external magnetic field at liquid helium temperatures.
The longitudinal nuclear spin relaxation rate varies exponentially with inverse temperature.

\end{abstract}

\maketitle
Recently, nuclear spins in semiconductors have attracted great attention due to the possible application for quantum information technology \cite{Kane1998,Bennett2000}.
While the nuclear Zeeman energy is extremely small even in the strong magnetic field and the polarization is negligible in thermal equilibrium, the nuclear spins can be dynamically polarized using the electron-nucleus flip-flop process arising from the contact hyperfine interaction.
The Hamiltonian is
\begin{eqnarray}
A {\mathbf I} \cdot {\mathbf S} = \frac{A}{2} [ I_+ S_- + I_- S_+] +AI_zS_z,
\end{eqnarray}
where $A$ ($>0$) is the hyperfine constant, and ${\mathbf I}$ and ${\mathbf S}$ are the nuclear spin and electron spin, respectively.
For GaAs, the effective electron $g$-factor is negative ($g = -0.44$) and the magnetic moment of electrons is parallel to the spin angular momentum as well as those of the lattice nuclei.
Electron spin resonance (ESR) measurements have shown that the relaxation of electron spin polarization via the first term causes the nuclear polarization and a shift of the ESR line (Overhauser shift) due to the second term \cite{Dobers1988,Berg1990}.

For local manipulation of dynamic nuclear polarization (DNP), electrical methods of inducing electron spin flip are desirable.
Previous works concerning the current-driven DNP can be classified into two categories: DNP induced by electron tunneling between spin-resolved edge channels in the integer quantum Hall regime \cite{Wald1994,Dixon1997,Machida2002,Machida2003,Wurtz2005} and DNP related to the transition between spin-polarized and spin-unpolarized fractional quantum Hall states \cite{Kronmuller1998,Kronmuller1999,Eom2000,Smet2002,Hashimoto2002,Kraus2002}.
Both of them require an extremely low temperature far below 1~K and most of the experiments were performed using dilution refrigerators.

In this letter, we report local manipulation of DNP using hot electrons in the integer quantum Hall regime.
We observe the resistance change due to DNP at liquid helium temperatures up to 4.2~K.
The longitudinal nuclear spin relaxation time $T_1$ is found to vary exponentially with inverse temperature.

Figure 1(a) shows a schematic diagram of DNP process induced by an electron temperature change, which will be demonstrated later.
At the Landau level filling factor $\nu={\rm odd}$, the chemical potential $\mu$ lies in the gap between spin-split levels and the electron spin polarization strongly depends on the electron temperature $T_e$, which can be easily controlled by the current density.
In the region where electrons are cooled (heated), electron spin flips from down to up (up to down) occur predominantly.
Part of electron spin flips involve the ``flop'' of nuclear spin via the contact hyperfine interaction conserving the total spin angular momentum, although most of them take place through other interactions such as spin-orbit interaction \cite{Zawadzki2004,Zutic2004}.
Thus a change in $T_e$ can cause DNP whose direction depends on whether electrons are cooled or heated.

\begin{figure}[t!]
\includegraphics[width=6cm]{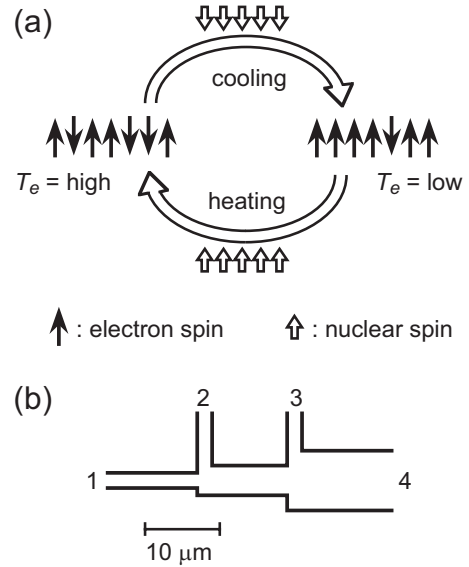}
\caption{
(a) Schematic diagram of dynamic nuclear polarization process caused by electron cooling and heating. (b) Schematic of the sample geometry.
}
\end{figure}

The sample is fabricated from a GaAs/AlGaAs heterostructure with an electron density of $4.8 \times 10^{15}~{\rm m}^{-2}$ and a mobility of $78~{\rm m}^2 / {\rm Vs}$ after brief illumination of a red LED.
It is mounted on a rotatory stage immersed in liquid ${}^4$He and the angle of the external magnetic field is fixed at 24 degrees away from the perpendicular to the interface.
The geometry of the sample is shown in Fig.~1(b).
The width $w$ of the current channel varies stepwise (2~$\mu$m, 4~$\mu$m, 8~$\mu$m).
Suppose a large current of electrons flows from the contact 1 to 4 ($I_{41} >0$).
Hot electrons injected from the narrow channel ($w=2~\mu{\rm m}$) are expected to be cooled in the central region ($w=4~\mu{\rm m}$) between the voltage probes 2 and 3 (Ref.~\onlinecite{Komori2005}) and to cause DNP against the external magnetic field.
Since $A$ in Eq.~(1) is positive, the negative DNP ($\langle I_z \rangle <0$) enhances the electron Zeeman energy owing to the second term.

Figure~2(a) shows the time evolution of four terminal resistance $R_{14,23}$ at $\nu=3$, which is measured using a small ac current $I_{14}^{\rm ac} \leq 0.1~\mu{\rm A}$ after applying dc current $I_{41}$ for 10 minutes.
For $I_{41}=+1~\mu{\rm A}$, a negative deviation of $R_{14,23}$ from the equilibrium value is observed.
It is attributed to the DNP induced enhancement of electron Zeeman energy and exhibits a slow exponential decay behavior.
A similar but opposite behavior observed after applying $I_{41}=-2~\mu{\rm A}$ shows that electron heating in the central region causes DNP toward the external magnetic field.
The nuclear origin of the resistance change is confirmed by means of nuclear magnetic resonance (NMR).
When the frequency of the oscillating magnetic field matches the resonance frequencies, the polarization of the nuclei quickly decreases and the deviation of $R_{14,23}$ from the equilibrium value becomes smaller.
The electrically detected NMR is observed for all the lattice nuclei ${}^{69}$Ga, ${}^{71}$Ga, ${}^{75}$As.
The full width at half maximum for ${}^{71}$Ga is obtained to be about 30~kHz, which is in the same order of magnitude as the values reported in the previous works \cite{Dixon1997,Machida2002,Kronmuller1999,Hashimoto2002}.
\begin{figure}[t!]
\includegraphics[width=7.5cm]{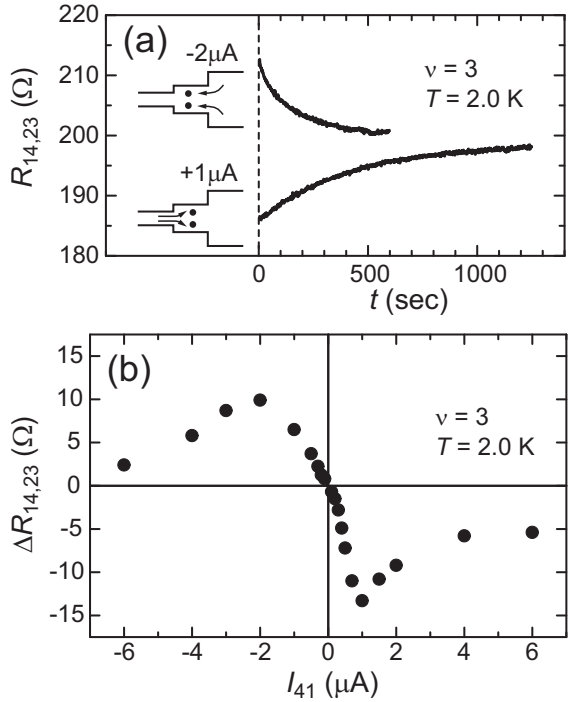}
\caption{
(a) Time evolution of $R_{14,23}$ after applying dc current $I_{41}=+1~\mu{\rm A}$ (lower) and $-2~\mu{\rm A}$ (upper) for 10~min.  (b) Deviation of $R_{14,23}$ from the equilibrium value at $t=0$ as a function of $I_{41}$ applied for $t<0$.
}
\end{figure}

The resistance change $\Delta R_{14,23}$ just after applying dc current ($t=0$) is shown in Fig.~2(b) as a function of $I_{41}$.
The magnitude of $\Delta R_{14,23}$ exhibits maxima for both directions of $I_{41}$ at $+ 1~\mu{\rm A}$ and $- 2~\mu{\rm A}$, respectively.
To make DNP efficiently in the central region \cite{Komori2005}, an electron temperature change from the current injection channel is essential.
For $I_{41}=+ 1~\mu{\rm A}$, the magnitude of the current density $j=|I_{41}/w|$ changes from 0.5 A/m to 0.25 A/m while it changes from 0.25 A/m to 0.5 A/m for $I_{41}=- 2~\mu{\rm A}$.
A nonlinear relationship between $V_{32}$ and $I_{41}$ indicates that the electron temperature (spin polarization) significantly changes in the range of $0.25~{\rm A/m} < j < 0.5~{\rm A/m}$.
In fact, the derivative $\partial (V_{32} / I_{41}) / \partial j$ has the maximum value at $j_{\rm MAX}=0.3$~A/m and  $V_{32} / I_{41}$ varies from 4~k$\Omega$ at $j=0.25$~A/m to 6~k$\Omega$ at $j=0.5$~A/m.
The critical current for the breakdown of the quantum Hall effect is evaluated to be $j_c \approx 0.5$~A/m from the magnetic field dependence of the critical Hall field for $\nu={\rm odd}$ at much lower temperature ($\sim 0.1~{\rm K}$) \cite{Kawaji1996}.
The reduction of $j_{\rm MAX}$ from $j_c$ can be attributed to the finite temperature effect.

Figure 3 shows the inverse of the relaxation time $T_1$ observed after applying dc current $I_{41}=+2 \mu{\rm A}$ for 10~min.
The obtained longitudinal nuclear spin relaxation rate $T_1{}^{-1}$ exhibits a deep minimum around $\nu=3$ (Fig.~3(a)).
Similar behavior was observed in the previous ESR measurement by using Overhauser shift \cite{Berg1990}.
Since the electron gyromagnetic ratio is much larger than the nuclear gyromagnetic ratio, the electron-nucleus flip-flop process is severely restricted by the energy conservation law \cite{Vagner1995}.
The minimum of $T_1{}^{-1}$ at $\nu={\rm odd}$ was explained by the enhancement of the electron Zeeman energy due to exchange interactions \cite{Berg1990,Vagner1995,Kim1994}.
\begin{figure}[t!]
\includegraphics[width=8cm]{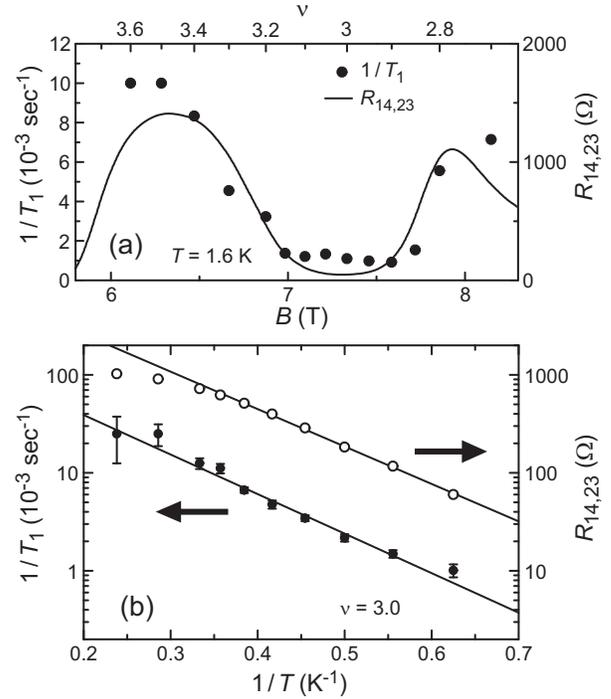}
\caption{
Longitudinal nuclear spin relaxation rate $1/T_{1}$ is shown together with the equilibrium value of $R_{14,23}$.
(a) Magnetic field dependence at $T=1.6~{\rm K}$.  (b) Temperature dependence at $\nu=3$.
}
\end{figure}

In order to clarify the mechanism of the relaxation process, we study the temperature dependence of $T_1{}^{-1}$.
$\Delta R_{14,23}$ can be clearly observed even at 4.2~K.
As shown in Fig.~3(b), $T_1{}^{-1}$ varies exponentially with inverse temperature.
The Arrhenius-type temperature dependence supports a phonon-assisted mechanism predicted by Kim, Vagner and Xing \cite{Kim1994}.
In this model, the energy conservation law is satisfied by absorbing a phonon whose energy is equal to the electron Zeeman energy.
The relaxation rate is proportional to the number of thermally excited phonons.
Using $T_1{}^{-1} \propto \exp[-E_z /T]$, we obtain the electron Zeeman energy $E_z = 9.3$~K from the experimental data.
Although this value is much larger than the bare Zeeman energy $|g|\mu_B B =2.2$~K, it is smaller than 17.6~K obtained from the temperature dependence of resistivity using $R_{14,23} \propto \exp[-E_z/2T]$.
At this stage, the role of the exchange interaction in the electron-nucleus flip-flop process is not clear.
Further investigations are required to understand the discrepancy between the enhancement factors of the electron Zeeman energy obtained from $T_1 {}^{-1} (T)$ and $R_{14,23} (T)$.

In summary, we have demonstrated dynamic nuclear polarization induced by an electron temperature change.
This method can be used at liquid helium temperatures and will be applied in various studies of nuclear spin dynamics in semiconductors.
Furthermore, we measured the longitudinal nuclear spin relaxation rate.
It was found to vary exponentially with inverse temperature.

We thank J. Matsunami for helpful discussions.
This work was partly supported by Sumitomo Foundation, Grant-in-Aid for Scientific Research (B) (No. 18340080) and Grant-in-Aid for Scientific Research on Priority Area "Physics of new quantum phases in superclean materials" (No. 18043008) from MEXT, Japan.

\clearpage

\end{document}